%% file: main.tex
\documentclass[lettersize,journal]{IEEEtran}
\usepackage{amsmath,amsfonts}
\usepackage{algorithmic}
\usepackage{algorithm}
\usepackage{array}
\usepackage[caption=false,font=normalsize,labelfont=up,textfont=up]{subfig}
\usepackage{textcomp}
\usepackage{stfloats}
\usepackage{url}
\usepackage{verbatim}
\usepackage{graphicx}
\usepackage{cite}
\usepackage{multirow}
\usepackage{subfig}
\usepackage{xcolor}
\usepackage{verbatim}
\usepackage{booktabs}
\usepackage{arydshln}
\usepackage{bbding}

\begin{document}

\title{Towards Better Domain Adaptation for Self-supervised Models: A Case Study of Child ASR}

\author{Ruchao Fan, 
        Yunzheng Zhu,
        Jinhan Wang, ~\IEEEmembership{Student Member, ~IEEE,} \\
        Abeer Alwan, ~\IEEEmembership{Fellow, ~IEEE}
\thanks{R. Fan, Y. Zhu, J. Wang are students in the ECE Department, University of California, Los Angeles,
CA, 90095, USA (e-mail: \{fanruchao,yunzhengzhu19,wang7875\}@g.ucla.edu).}
\thanks{A. Alwan is a Professor in the ECE Department, University of California, Los Angeles,
CA, 90095, USA (e-mail: alwan@ee.ucla.edu)}}



\maketitle

\begin{abstract}
Recently, self-supervised learning (SSL) from unlabelled speech data has gained increased attention in the automatic speech recognition (ASR) community. Typical SSL methods include autoregressive predictive coding (APC), Wav2vec2.0, and hidden unit BERT (HuBERT). However, SSL models are biased to the pretraining data. When SSL models are finetuned with data from another domain, domain shifting occurs and might cause limited knowledge transfer for downstream tasks. In this paper, we propose a novel framework, domain responsible adaptation and finetuning (DRAFT), to reduce domain shifting in pretrained speech models, and evaluate it for a causal and non-causal transformer. For the causal transformer, an extension of APC (E-APC) is proposed to learn richer information from unlabelled data by using multiple temporally-shifted sequences to perform prediction. For the non-causal transformer, various solutions for using the bidirectional APC (Bi-APC) are investigated. In addition, the DRAFT framework is examined for Wav2vec2.0 and HuBERT methods, which use non-causal transformers as the backbone. The experiments are conducted on child ASR (using the OGI and MyST databases) using SSL models trained with unlabelled adult speech data from Librispeech. The relative WER improvements of up to 19.7\% on the two child tasks are observed when compared to the pretrained models without adaptation. With the proposed methods (E-APC and DRAFT), the relative WER improvements are even larger (30\% and 19\% on the OGI and MyST data, respectively) when compared to the models without using pretraining methods.

\end{abstract}
\begin{IEEEkeywords}
self-supervised learning, end-to-end speech recognition, children's ASR, domain adaptation, residual adapters
\end{IEEEkeywords}

\input{Tex/introduction}
\input{Tex/ssl_general}

\input{Tex/method}

\input{Tex/exp_steup}

\input{Tex/result}

\input{Tex/conclusion}

\section{Acknowledgement}
This paper was supported in part by the NSF and the UCLA-Amazon Science Hub. Thanks to PAII Inc. for offering their GPU platforms for some experiments.

\bibliographystyle{IEEEtran}
\bibliography{ref}

\vfill

\end{document}

%% file: Tex/introduction.tex
\section{Introduction}
\IEEEPARstart{D}{espite} impressive advancement in developing automatic speech recognition (ASR) techniques in the last decade, children's ASR remains difficult. Challenges arise, in part, from difficulties in acoustic and language modeling of child speech. Due to different growth patterns of children and motor control issues, child speech has a higher degree of intra-speaker and inter-speaker acoustic variability than adult speech~\cite{lee1999acoustics}. Additionally, child speech is characterized by significant mispronunciations and disfluencies~\cite{yaruss1999language,tran2020analysis}. Another challenge is the lack of large-scale publicly-available child speech databases, and thus child ASR can be treated as a low-resource task\cite{wang2021low}.


Recently, self-supervised learning (SSL) from speech data has been investigated\cite{chen2021wavlm,zhang2021bigssl,pmlr-v139-wang21y,chung2021w2v,ao2021speecht5,jiang2021further,wang2020unsupervised,liu2020mockingjay,liu2021tera} because of its great potential of improving low-resource tasks through learning prior knowledge from large amounts of data without annotations. SSL models can be used in two manners: 1) feature extraction to replace human-designed features\cite{yang2021superb,evain2021lebenchmark,chang2021exploration}; and 2) model initialization for finetuning downstream tasks\cite{vyas2021comparing,misra2021comparison}. The idea of SSL is to design pseudo-labels for training deep neural networks (DNN) and then transfer the learned knowledge to a downstream supervised task. For example, autoregressive predictive coding (APC) uses temporally-shifted sequences to perform prediction such that the model predicts future frames from previous frames\cite{chung2019unsupervised,chung2020generative,Ravi2020}. In \cite{fan2021bi}, we proposed a bidirectional APC (Bi-APC) method for bidirectional long short-term memory (BLSTM) pretraining for children's ASR. Different from APC and Bi-APC where the reconstruction loss is used, Wav2vec-based methods are implemented to include negative samples, and a contrastive loss is utilized to increase the distance from the output to negative samples and decrease that distance to the positive sample\cite{oord2018representation,schneider2019wav2vec,baevski2020wav2vec,baevski2019vq}. The positive sample is the frame being masked (to be predicted), and negative samples are the unmasked frames in the utterance.
A more recent SSL framework, HuBERT\cite{hsu2021hubertc,hsu2021hubertj}, creates the pseudo-label of each speech frame using clustering techniques like K-means. These methods have been shown to be effective for low-resource ASR tasks such as low-resource languages\cite{riviere2020unsupervised,yi2020applying}, noisy speech\cite{wang2021wav2vec} and accented speech\cite{li2021accent}. 


However, a weakness of SSL training is domain shifting that happens when the domain of the finetuning data is different than that of the pretraining data\cite{meng2021don,hsu2021robust}. Although a performance improvement can be observed when the magnitude of the pretraining data is large enough, previous work has shown that additional gains can be obtained by including target domain data in the ASR pretraining stage\cite{hsu2021robust,hwang2021large}. But including target domain data would be impractical if we are not aware of the finetuning task at the pretraining stage. In addition, retraining a large-scale SSL model with both the source and target domain data to address domain shifting may not always be possible or computationally efficient. Hence, investigating adaptation methods for SSL is gaining attention for work involving out-of-domain low-resource tasks. Previous studies proposed to perform adaptation of supervised models either during or after the finetuning stage \cite{khurana2021magic,huo2021incremental}. No additional adaptation stage of self-supervised models has been investigated before for domain shifting in SSL methods. 

In this paper, building on our work in \cite{fan2021bi}, we explore how SSL methods can improve the performance of child ASR in the context of a low-resource setting for causal and non-causal transformers. First, autoregressive predictions at different temporal distances are shown to enable the pretrained model to learn more effectively \cite{chung2020generative}. We, therefore, propose to use multiple temporally-shifted sequences to construct a multi-task training objective for APC. Second, the proposed Bi-APC framework in our previous work performs well for BLSTM, whose parameters can be separated into forward-related and reverse-related ones. It is unknown whether the Bi-APC framework can be used for transformer architectures that have only one set of parameters. To do so, we copy the modules in the transformer during pretraining and treat the modules as separate parameters for two APCs in two directions. After pretraining, we either use one of the modules when the weights are shared, or average the weights of the two modules to formalize the final parameters for finetuning. Finally, we propose a domain responsible adaptation and finetuning (DRAFT) framework to address the domain shifting problem in SSL. In DRAFT, residual adapters are placed between blocks in the transformer and are responsible for learning domain related information at an additional adaptation stage. The additional adaptation stage trains the model with target finetuning data and with the same SSL loss that was used in the pretraining stage. Only residual adapters are updated during the adaptation stage so that the knowledge learned from source domain data can be retained. The proposed DRAFT framework has a lower cost than adding target domain data at the pretraining stage and can be used in various SSL methods.

Note that residual adapters have been proposed before in the literature. In \cite{tomanek2021residual, houlsby2019parameter}, residual adapters are inserted to achieve a parameter efficient adaptation for low-resource supervised tasks, but the performance is worse than finetuning the entire model. The method is beneficial when adaptation is frequently required such as personalization of a speech recognition model. In \cite{hou2021exploiting,kannan2019large,rebuffi2017learning}, residual adapters are applied to learn domain specific parameters to achieve robust models for various domains. Differences between our work and these methods will be discussed further in Sec.\ref{sssec:res_adapt}. The DRAFT part is an extension of our recent paper \cite{fan2022draft}. We report on more experiments in this paper to better understand the functionality of the residual adapters.

The contributions of this paper are:

\begin{itemize}
    \item{An extension to autoregressive predictive coding (E-APC) is proposed so that the pretrained model can learn more useful speech representations from unlabelled data. It is then used for a causal transformer pretraining.}
    \item{Various solutions for using the Bi-APC algorithm in non-causal transformers are investigated.}
    \item{A \textbf{d}omain \textbf{r}esponsible \textbf{a}daptation and \textbf{f}ine\textbf{t}uning (DRAFT) framework is proposed to address the domain shifting problem in self-supervised pretrained models. Different from \cite{fan2022draft}, DRAFT's performance is examined with Bi-APC, and ablation studies are conducted for a better understanding of the DRAFT framework.}
\end{itemize}

The remainder of this paper is organized as follows. Section \ref{sec:ssl_general} proposes a general SSL framework. Section \ref{sec:method} describes the proposed methods for better improving low-resource ASR tasks with SSL pretrained models. Experimental setups are described in Section \ref{sec:exp_setup}. Results are shown and discussed in Section \ref{sec:result}. We conclude the paper in Section \ref{sec:conclusion}.

%% file: Tex/ssl_general.tex
\section{A General Self-supervised Learning Framework}
\label{sec:ssl_general}
\begin{figure}[t]
\centering
\centerline{\includegraphics[width=0.32\textwidth,height=0.37\textwidth]{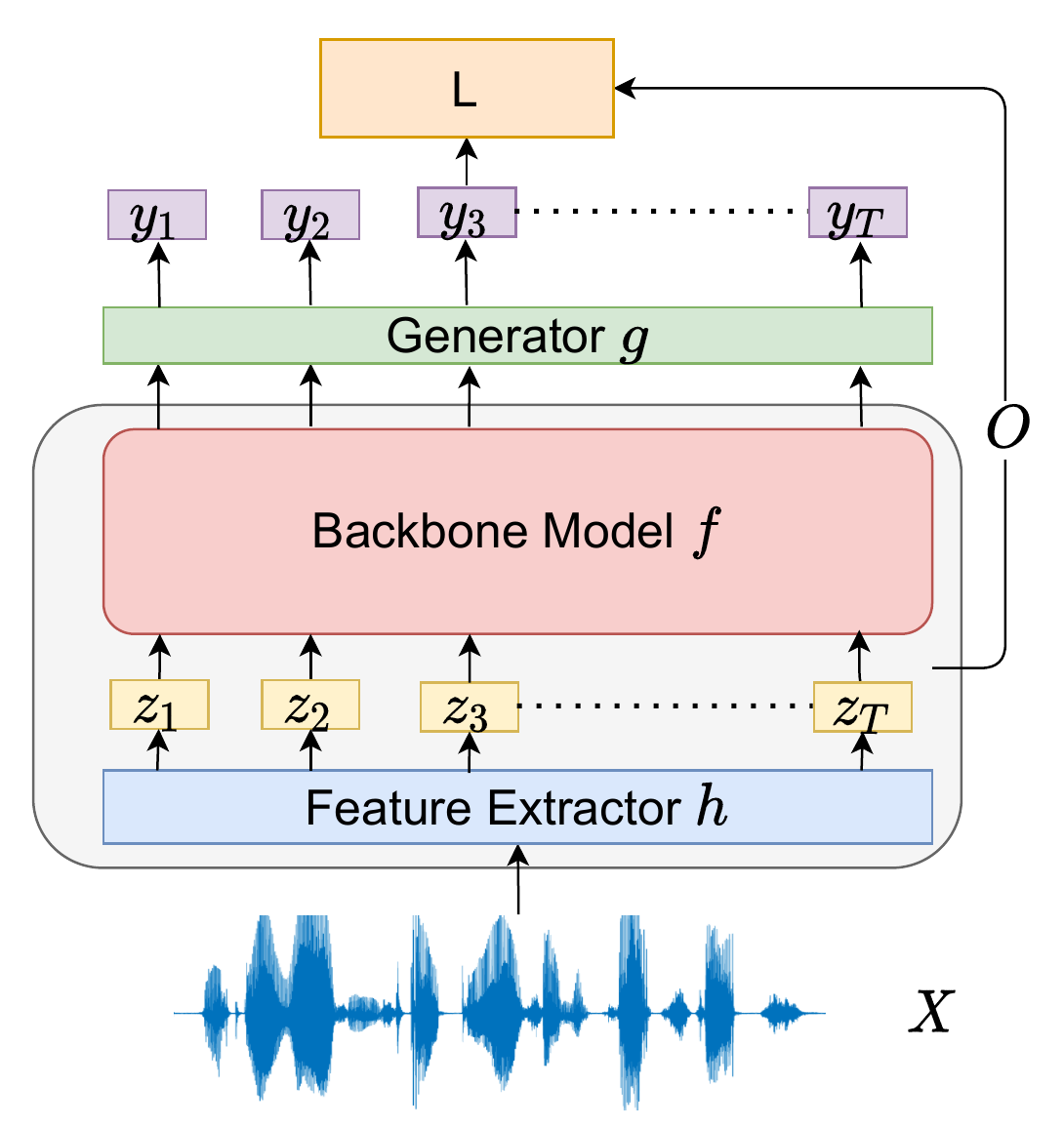}}
\caption{Proposed self-supervised learning framework. $h$ is a function to extract speech representation $Z$. $O$ is an operation over $Z$. $f$ is the backbone model for pretraining. $g$ is a generator that maps the output of the backbone model to have the same dimension as $O(Z)$ and outputs $Y$. $L$ computes the SSL loss using $Y$ and $O(Z)$.}
\label{fig:ssl_general}
\end{figure}

Self-supervised learning (SSL) learns useful speech representations for downstream tasks without explicit supervision. After pretraining, the model can be used for model initialization for downstream tasks. We propose a general framework for various SSL methods, which is illustrated in Fig.\ref{fig:ssl_general}. 

Let $X=(x_1,...,x_i,...,x_n)$ denote the raw waveform of an utterance, where each $x_i$ is a sampled data point. The self-supervised learning methods first extract representations $Z=(z_1,...,z_t,...,z_T)$ for each frame $t$ using a function $h$, which in general can be either a human-designed function, like an MFCC extractor\cite{harris1978use}, or a learned deep network, like a convolution neural network\cite{lecun1995convolutional}. There may be a special case when $X$ represents human-designed spectral features. Then $h$ is a stack of a human-designed function and the module that maps the spectral features to latent representations for prediction. A backbone model $f$ parameterized with $\theta$ is then used to build contextualized representations. A generator $g$ finally converts the contextualized representations into a prediction space with a pre-defined dimension and outputs $Y=(y_1,...,y_t,...,y_T)$. An operation $O$ over speech representation $Z$ is designed to obtain a pseudo-label for the task. The key idea of SSL is to construct a loss function $L$ between $O(Z)$ and model output $Y$, ensuring that no information leaking appears in the forward computation so that trivial solutions are ignored during the optimization process. As a result, the SSL objective function is:
\begin{equation}
  L_{\text{SSL}} = L(f(h(X)), O(h(X)))
  \label{eq:sslloss}
\end{equation}
We omit $g$ for simplicity because it can be regarded as a part of $f$. The main differences between various SSL methods are the definition of $O$ for obtaining a supervision and of $L$ as an optimization objective.

In this section, we discuss how several SSL methods can be used in our proposed framework. These methods include autoregressive predictive coding\cite{chung2019unsupervised,chung2020generative}, Wav2vec2.0\cite{baevski2020wav2vec} and hidden unit BERT (HuBERT)\cite{hsu2021hubertc,hsu2021hubertj}.

\subsection{Autoregressive Predictive Coding}
\label{ssec:bg_apc}
Autoregressive predictive coding (APC) uses human-designed features as model input ($Z=h(X)$). Typically, 80-dimensional log-mel filter-bank features are used as $Z$. APC utilizes a temporally-shifted sequence to predict the frame $n$ steps ahead of the current frame given all previous frames. As a consequence, the operation $O$ with a temporal lag of $n$ up to the time step $T-n$ satisfies $O_n(\{z_1,z_2,...,z_{T-n}\})=\{z_{1+n},z_{2+n},...,z_{T}\}$. Since $Z$ are not latent representations in APC, the $L_p$ norm distance can be used as the loss function. The final objective function is formulated as follows:
\begin{equation}
  L_{\text{APC}} = L_p(f(Z), O_n(Z)) = \sum_{t=1}^{T-n}(|y_t-z_{t+n}|_p)
  \label{eq:apcloss}
\end{equation}
where $n$ is fixed as a hyper-parameter. APC essentially adopts neural language model style training using speech features instead of word embeddings. The mechanism is suitable for online speech recognition model pretraining because APC considers information from only one direction. It is, however, not suitable for bidirectional model pretraining. In this paper, we conduct experiments to explore whether APC can be extended to bidirectional model pretraining.

\subsection{Wav2vec2.0}
\label{ssec:bg_wav2vec}
Wav2vec2.0 has evolved from contrastive predictive coding (CPC)\cite{oord2018representation}, Wav2vec\cite{schneider2019wav2vec}, and Vq-Wav2vec\cite{baevski2019vq,baevski2019effectiveness}. We only study Wav2vec2.0 because of its better performance.

Wav2vec2.0 uses raw waveforms as model inputs, which means $h$ is a parameterized model for learning feature extraction. $h$ consists of multiple blocks of temporal convolution layers with a total stride that decreases the sequence length from the number of sampled points to the number of frames. Different from APC, Wav2vec2.0 adopts masked language model (MLM)\cite{devlin2018bert} style training, where the backbone model $f$ tries to reconstruct masked speech representations. Let $M$ be the mask operation onto speech representation $Z$. We define $Z_1$ as the original tokens that will be masked by $M$, and $Z_2$ as the original tokens that will not be masked. Then, when applying $M$ to $Z$, we obtain $Z_{mask}$ and $Z_{obs}$ as the masked and unmasked tokens. The corresponding outputs are referred to as $Y_{mask}$ and $Y_{obs}$. Only $Y_{mask}$ contributes to the loss computation. Hence, the forward computation of $f$ could be formulated as $Y_{mask}=f(Z_{mask} \oplus Z_{obs}) - Y_{obs}$. Suppose the length of the masked proportion is $U$, we write $Y_{mask}$ as $\{y_{mask}^1, y_{mask}^2, ...,y_{mask}^U\}$.  Since $Z$ are latent representations, the contrastive loss is preferred so that the true latent is distinguished from distractors. A vector-quantization (vq) layer is also inserted after $Z$ to obtain more compact representations for supervision so that the model can learn more efficiently. The operation $O$ can be summarized as  $O(Z)=\text{vq}(Z_1) \oplus \text{Sample}(\text{vq}(Z_2))=\{(q_{pos}^1, Q_{neg}^1), (q_{pos}^1, Q_{neg}^1),...,(q_{pos}^U, Q_{neg}^U)\}$, where $q_{pos}$ is a positive sample after vq layers, and $Q_{neg}$ is a set for negative samples as distractors in the contrastive loss. The objective function is formulated as:

\begin{equation}
\label{eq:wav2vecloss}
\begin{aligned}
  L_{\text{wav2vec2.0}} &= L_{ctras}(Y_{mask}, O(Z)) \\
    &= -\sum_u log \frac{\exp(sim(y_{mask}^u, q_{pos}^u))}{\sum_{q_{neg}^u\in Q_{neg}^u} \exp(sim(y_{mask}^u, q_{neg}^u))} \\
\end{aligned}
\end{equation}
where $sim(a,b)$ is the cosine similarity between context representation $Y_{mask}$ and quantized latent representations $O(Z)$. There is also an additional diversity loss in Wav2vec2.0. Since the observed sequence has information from both directions for most masked frames, Wav2vec2.0 is suitable for bidirectional model pretraining. However, Wav2vec2.0 always requires more training iterations than APC \cite{hwang2022large}. This is because only a portion of the frames are masked for prediction in Wav2vec2.0 and the mask regions are different each time the sequence is trained.

\subsection{HuBERT}
\label{ssec:bg_hubert}
Hidden unit BERT (HuBERT) uses the same masked language model style training as Wav2vec2.0. Differently, HuBERT does not require negative samples. Instead, it introduces an acoustic unit discovery process before the pretraining stage. For example, the most useful strategy in HuBERT\cite{hsu2021hubertj} is performing K-means on MFCC features or intermediate model outputs to obtain a pseudo-label category for each frame. HuBERT creates a learned embedding for each category. The embedding of the true category is equivalent to the positive sample in Wav2vec2.0 and all other embeddings are essentially negative samples. Thus, the operation $O$ is an unit discovery process for HuBERT. The loss computation is similar to the fine-tuning task. If we define $O(Z)=(c_1, c_2,...,c_T)$, where $c_t$ is the pseudo-category for each frame, the objective function is computed as a weighted sum of both the masked and observed output sequences.  

\begin{equation}
\label{eq:hubertloss}
\begin{aligned}
  L_{\text{HuBERT}} &= L(Y, O(Z))  \\
    &= \alpha L(Y_{mask}, O(Z_{mask})) + (1-\alpha) L(Y_{obs}, O(Z_{obs}))\\
    &= -\alpha \sum_{c_t \in O(Z_{mask})} \log P(c_t | Z) - \\ & (1-\alpha) \sum_{c_t\in O(Z_{obs})} \log P(c_t | Z)\\
\end{aligned}
\end{equation}
where $\alpha$ is the task ratio. Similar to\cite{hsu2021hubertc,hsu2021hubertj}, we use $\alpha=1$ because we directly use the open-sourced pretrained HuBERT model\cite{ott2019fairseq} as initialization for children's ASR training. More importantly, we apply the proposed domain adaptation technique on these models to show its general effectiveness.

%% file: Tex/method.tex
\section{Methods}
\label{sec:method}

\begin{figure*}[ht]
\centering
\subfloat[No Sharing]{\includegraphics[width=0.22\textwidth,height=0.22\textwidth]{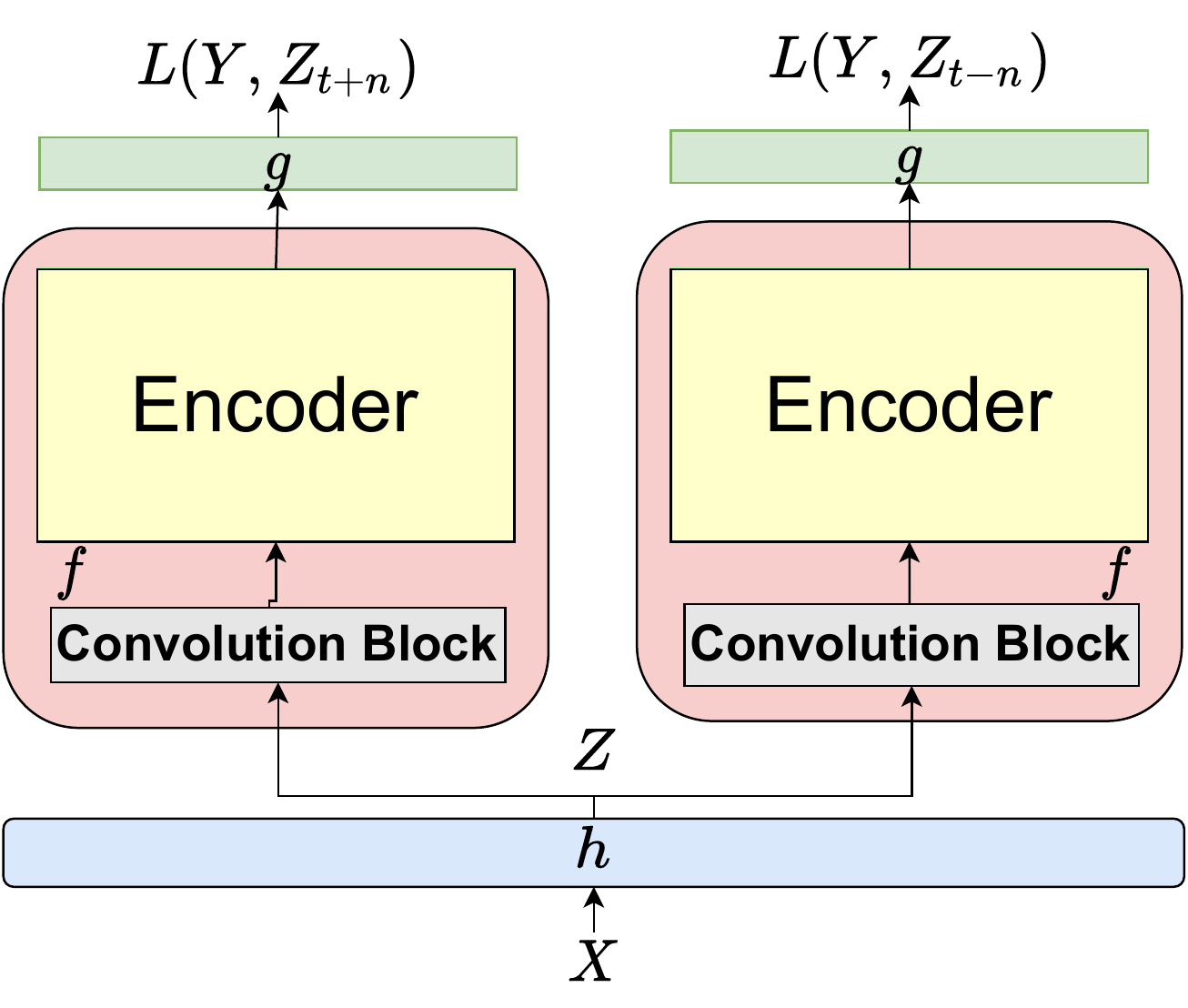}} %
\quad \quad
\subfloat[Share Generator $g$]{\includegraphics[width=0.22\textwidth,height=0.22\textwidth]{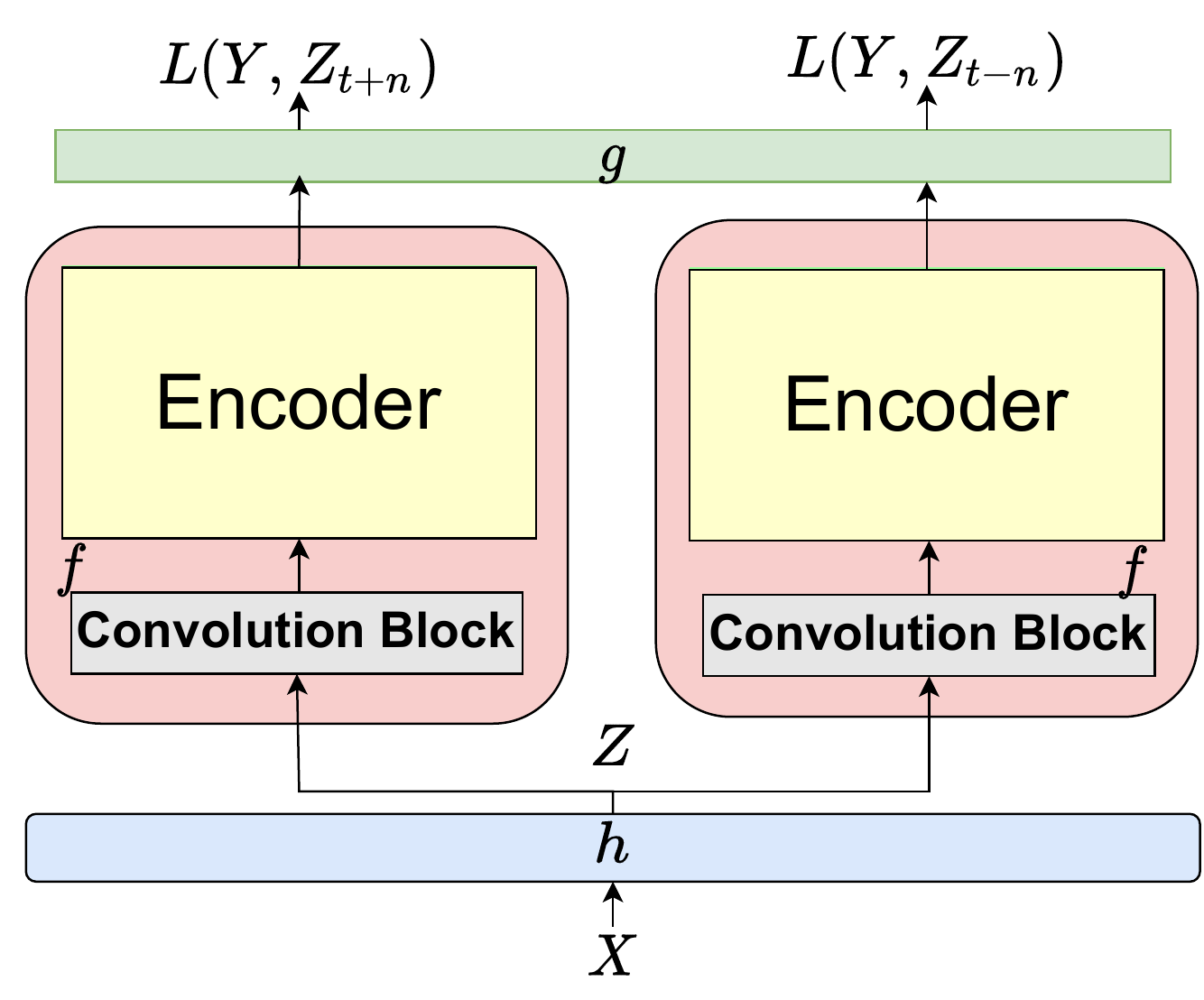}}
\quad \quad
\subfloat[Share $g$ and Encoder]{\includegraphics[width=0.22\textwidth,height=0.22\textwidth]{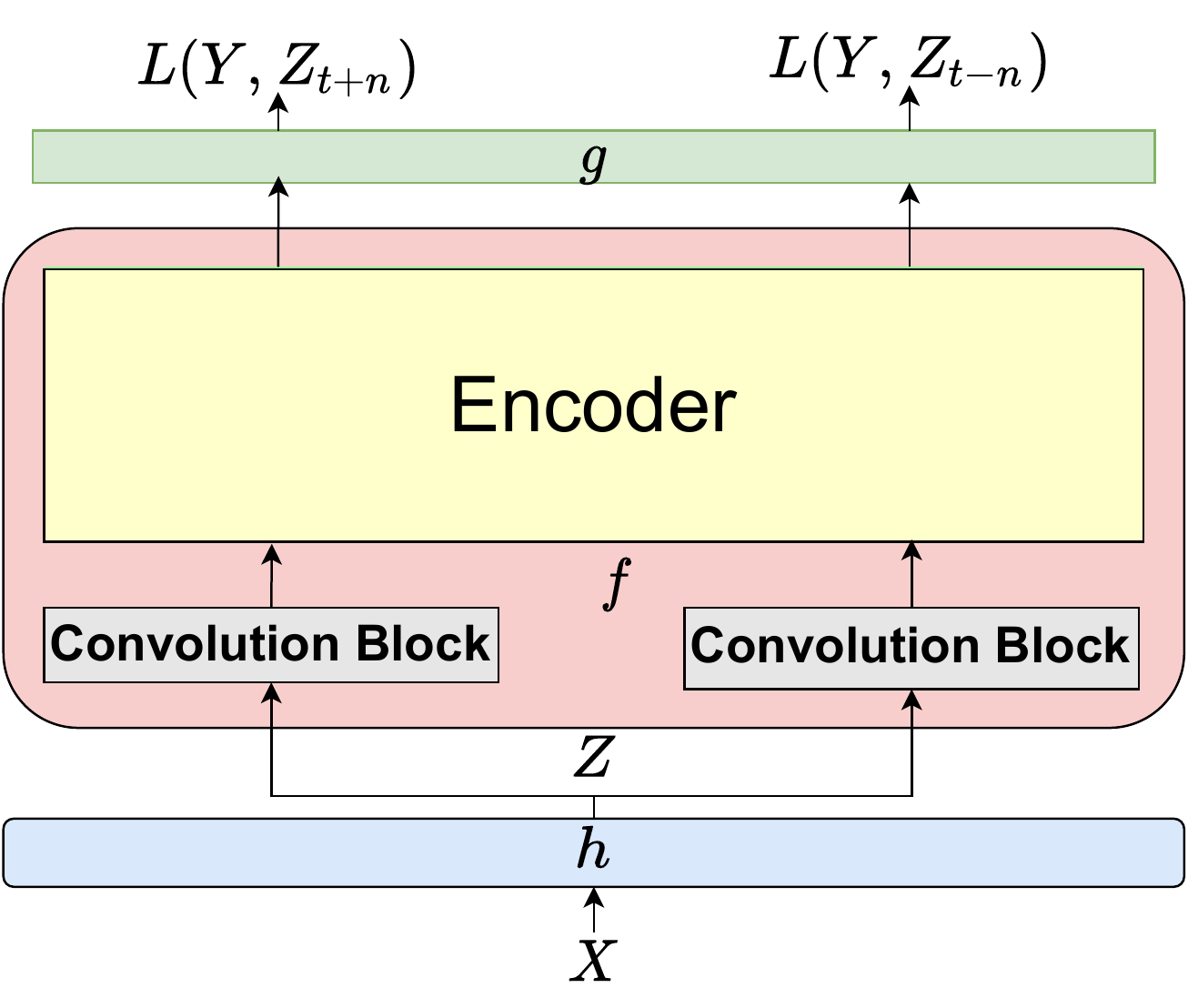}}
\quad \quad
\subfloat[Share All]{\includegraphics[width=0.22\textwidth,height=0.22\textwidth]{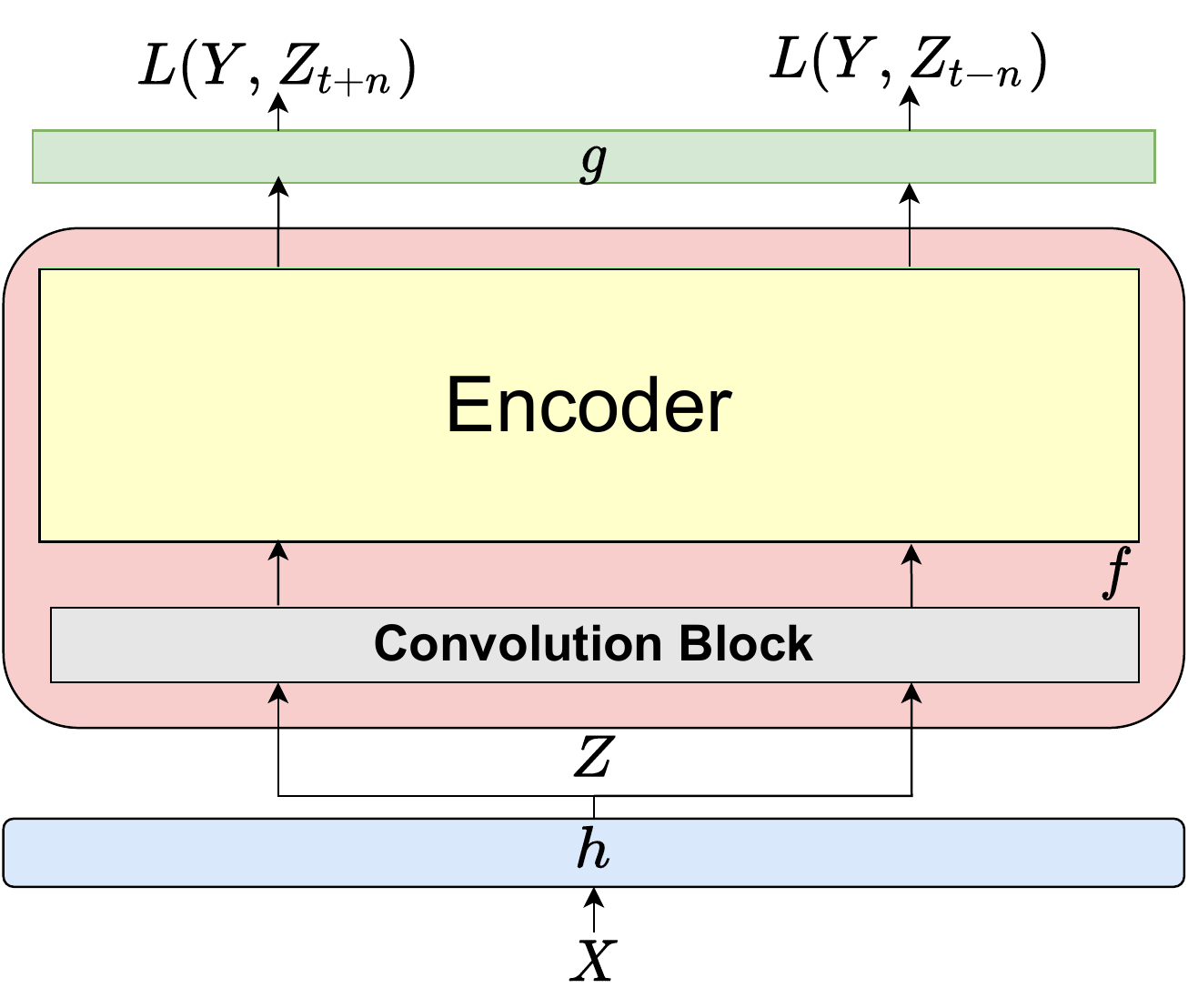}}
\caption{Various solutions for training a non-causal transformer with Bi-APC. Each color represents one module. Notations and blocks are consistent with those described in Section \ref{sec:ssl_general}.}
\label{fig:biapc_trans}
\end{figure*}

In this section, we first introduce an extension of autoregressive predictive coding (APC) for causal transformer pretraining. Then, bidirectional APC (Bi-APC) for non-causal transformer pretraining is elaborated on as an extension of our previous Bi-APC paper\cite{fan2021bi}. We end this section with DRAFT, the proposed adaptation framework for self-supervised pretrained models.

\subsection{An Extension of APC}
\label{ssec:apc_extend}
The original APC\cite{chung2019unsupervised} uses one temporally-shifted sequence $Z_{t+n}$ during pretraining, as shown in Eq.\ref{eq:apcloss}. A model may learn differently with different temporal lags, aka. different values of $n$. For example, the model learns to exploit local smoothness of the signal with a small value of $n$, while it learns a global structure with a large value of $n$. Hence, it is intuitive to include multiple temporally shifted sequences with different lags during pretraining and reformulate APC as a multi-task training loss. If we regard Eq.\ref{eq:apcloss} as $L_{\text{APC}}^n$, the extension of APC (E-APC) has the following objective function:
\begin{equation}
  L_{\text{E-APC}} = \sum_{n=s}^{s+k} L_{\text{APC}}^n= \sum_{n=s}^{s+k} \sum_{t=1}^{T-n}(|y_t-z_{t+n}|_p)
  \label{eq:eapcloss}
\end{equation}
where $s$ is the temporally-shifted sequence with the smallest value of $n$ and $k$ is the number of consecutive temporally-shifted sequences used in pretraining. During implementation, the backbone model $f$ is shared across different tasks while each task has its own generator $g$.  

A recent paper also considered multiple targets for better APC pretraining\cite{chung2020improved}. An auxiliary predictive loss with the same temporal lag as the original APC loss is proposed based on an additional RNN for regularization. In our work, however, only one model (an RNN or transformer) is used to learn various temporal lags.

\subsection{Bi-APC for non-causal transformers}
\label{sssec:tsfm_biapc}
APC and E-APC are suitable for pretraining causal transformers because of their similar autoregressive mechanism that predicts future frames from previous frames. Since bidirectional models outperform their unidirectional couterparts\cite{zeyer2017comprehensive,6707742}, we consider the usage of APC for bidirectional model pretraining. In our previous work\cite{fan2021bi}, we successfully used APC for bidirectional long short-term memory (BLSTM) and proposed a Bi-APC framework. However, it is unknown whether Bi-APC can be applied to non-causal transformers that are bidirectional models with a transformer backbone. In this section, we discuss solutions for applying Bi-APC to non-causal transformers.

The parameters of BLSTM are designed to be separated into a left-to-right and right-to-left context modelling LSTMs. The Bi-APC framework takes each LSTM as an individual APC and ignores the parameters that induce information exchange between two LSTMs. When the extension of APC is also applied to Bi-APC, we can write the objective function of the Bi-APC as follows:
\begin{equation}
\begin{aligned}
  L_{\text{E-BiAPC}} &= \sum_{n=s}^{s+k} L_{\text{APC}}^{n} + \sum_{n=s}^{s+k} L_{\text{APC}}^{-n} \\
  &= \sum_{n=s}^{s+k} \sum_{t=1}^{T-n}(|y_t-z_{t+n}|_p) + \sum_{n=s}^{s+k} \sum_{t=n+1}^{T}(|y_t-z_{t-n}|_p) \\
\end{aligned}
\label{eq:ebiapcloss}
\end{equation}

The parameters in non-causal transformers, however, are not separated for contextual modelling from both directions. It is unknown whether the parameters would confuse the learning of individual APC loss in two directions. There are three major modules in a pretrained model: convolution block, encoder, and generator. We can essentially assume each module has two copies and train one causal transformer using a left-to-right APC and the other causal transformer using a right-to-left APC, like BLSTM. The trained modules are then averaged to be the final model initialization for the finetuning task. Parameters of the two copies could also be shared so that averaging is not needed after pretraining. As a result, we explore four various Bi-APC pretraining schemes for non-causal transformers as shown in Fig.\ref{fig:biapc_trans}: 1) no modules are shared; 2) only the generator is shared; 3) only the convolution block is not shared and 4) all modules are shared, which is similar to \cite{chen2020transformer}. The four solutions are selected based on the number of modules shared from top to bottom during Bi-APC pretraining. By exploring the four Bi-APC pretraining schemes, we can understand how the shared parameters affect APC pretraining in two directions and whether Bi-APC framework is suitable for models with shared parameters for bidirectional contextual modelling. Note that we use causal convolution layers in the convolution block and a causal mask in each self-attention layer in the encoder.

Another way of incorporating bidirectional contextual information from pretraining could be averaging or concatenating outputs from two pretrained causal transformers. However, this has been previously investigated in \cite{ling2020deep}. In addition, averaging or concatenating outputs will result in doubling model parameters with the goal of finding good speech representations for downstream tasks. In our case, however, we aim to find a good initialization of non-causal transformers for finetuning.

\subsection{DRAFT: Adaptation of Self-supervised Pretrained Models}
\label{ssec:ra_da}

Including data from the target domain in the pretraining stage can improve the performance of the target task. But this method requires knowledge of the target domain and then re-training the self-supervised model with a larger amount of data, which is time-consuming and computationally expensive. It would be more practical to adapt the pretrained models with target data only when the target domain is unknown in the pretraining stage. In this section, we propose DRAFT, a domain responsible adaptation and finetuning framework, to alleviate domain shifting in the conventional self-supervised pretraining and finetuning paradigm. DRAFT is a three-stage training paradigm with residual adapters inserted in the backbone model $f$. The residual adapters are designed to learn knowledge from the target domain data.

\subsubsection{Simple Adaptation for Finetuning (SAFT)}
Before introducing DRAFT, a simpler way of doing adaptation is to retrain the model with target data. An adaptation stage is inserted between the pretraining and finetuning stage. The adaptation stage continues to train the model from the pretraining stage with the same self-supervised loss function but with target data only. All the parameters in the model are updated at the adaptation stage. The model after the adaptation stage is used as initialization for the finetuning stage with an ASR loss.

\subsubsection{Domain Responsible Adapters for Finetuning (DRAFT)}
\label{sssec:res_adapt}

\begin{figure}[t]
\centering
\centerline{\includegraphics[width=0.38\textwidth,height=0.40\textwidth]{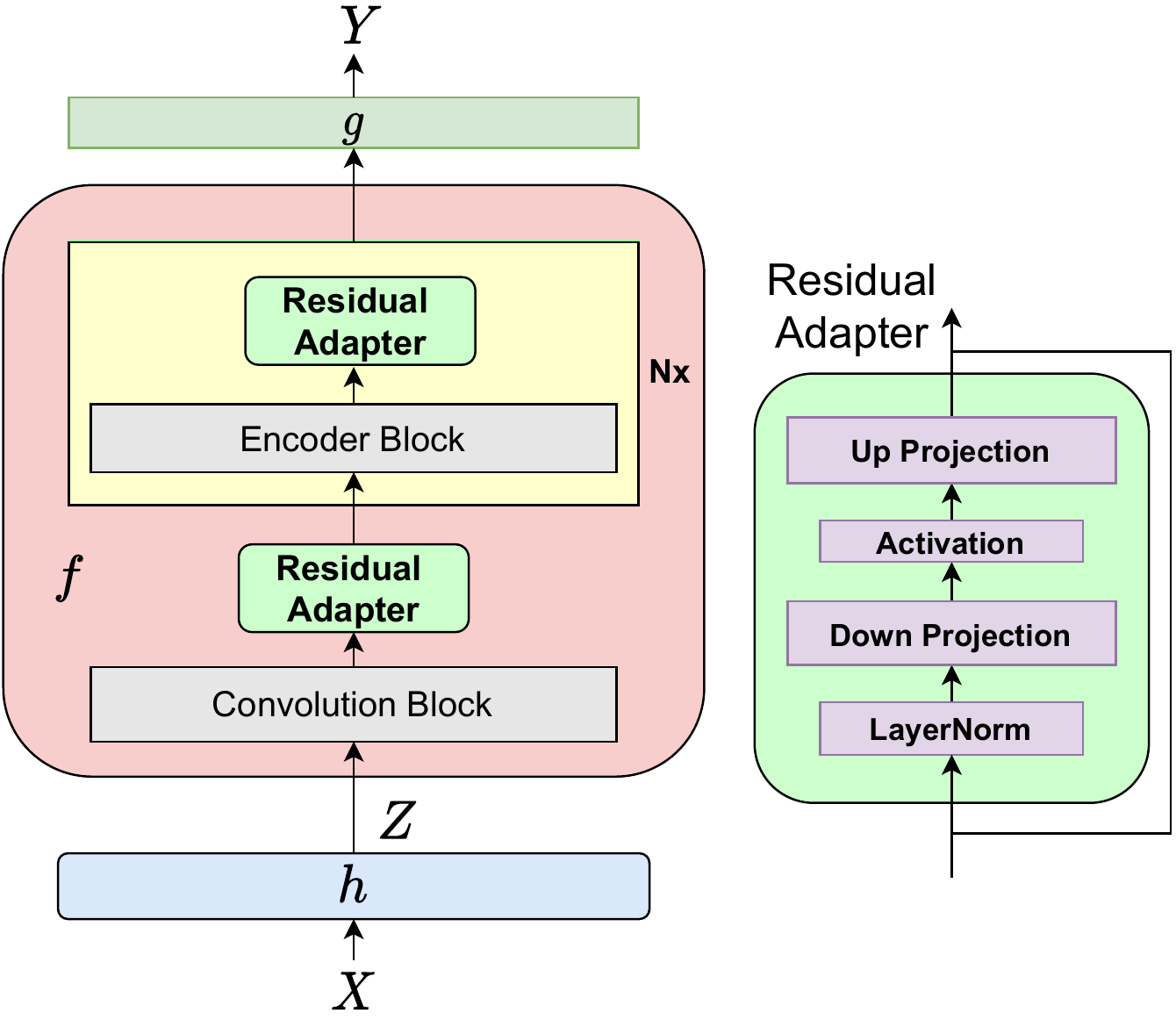}}
\caption{Structure of the backbone model $f$ with residual adapters inserted after each block. $N$x indicates that the module can be repeated $N$ times (proportional to the number of encoder blocks). $X, h, Z, f, g$ and $Y$ are the same as those in Fig.\ref{fig:ssl_general}. The right side of the figure shows the components in a residual adapter.}
\label{fig:res_adapt}
\end{figure}

SAFT updates the parameters of the entire model, and thus overfitting might occur, leading to a catastrophic forgetting for the self-supervised model\cite{french1999catastrophic,kessler2021continual,chang2021towards}. Knowledge learned at the pretraining stage may diminish because of an aggressive learning strategy. Not only does the domain shifting problem remain unsolved, but also a potential improvement breakdown may happen. To address this issue, we propose domain responsible adapters for finetuning (DRAFT) framework that uses residual adapters in the backbone model to learn from target domain data while retaining the source domain knowledge.


We show the backbone model with residual adapters in Fig.\ref{fig:res_adapt}. As shown in the figure, a residual adapter consists of two feed-forward layers with a layer normalization and a residual connection. The activation function between the two feed-forward layers makes the adapter non-linear. The number of parameters in the residual adapter depends on the dimension of the output after the down projection layer, which is defined as $d_{ada}$. The effect of $d_{ada}$ on performance will be explored experimentally. Note that the residual adapter can be placed anywhere in the model. In our case, we insert one residual adapter after the convolution block and one after each encoder block. We assume that the output of each block needs to be transformed to be similar to that of the target domain data so that the model can converge more easily. 

Residual adapters have been previously used for domain adaptation of supervised models\cite{tomanek2021residual, houlsby2019parameter}. However, we develop a way of using residual adapters for adaptation of self-supervised models via a three-stage training paradigm. The motivation is to prevent catastrophic forgetting that happens when finetuning the entire pretrained model, and to address the domain shifting problem in self-supervised learning. We also do not learn different residual adapters for different domains. Our goal is to find a better model initialization of the downstream low-resource tasks. In \cite{DBLP:conf/rep4nlp/KimSSH21}, residual adapters are used to re-pretrain and finetune the target domain data with the purpose of parameter efficiency in a natural language processing application. Hence, only residual adapters are updated at the finetuning stage, while we update the entire model, and the adaptation stage uses only finetuning data.

To better understand the algorithm, we detail the novel three training stages of DRAFT. Let $\theta_{ada}$ be the parameters in residual adapters, $\theta_f$ the parameters in the backbone model (without residual adapters), $\theta_g$ the parameters in the generator for the self-supervised task, and $\theta_g'$ the parameters in the generator for the ASR task. Suppose source domain data are $S_{src}$ and target domain data are $S_{tgt}$, the three-stage training paradigm can be described as:
\begin{itemize}
    \item Stage 1: Initialize a model $\{\theta_{f}^0, \theta_{g}^0\}$, update the parameters using data $S_{src}$ and self-supervised loss $L_{ssl}$, and obtain a pretrained model $\{\theta_{f}^1, \theta_{g}^1\}$.
    \item Stage 2: From model $\{\theta_{f}^1, \theta_{g}^1\}$, insert residual adapters after each block initialized with $\theta_{ada}^0$, freeze $\{\theta_{f}^1, \theta_{g}^1\}$ and update $\theta_{ada}^0$ using data $S_{tgt}$ and the same self-supervised loss $L_{ssl}$, and obtain an adapted model $\{\theta_{f}^1, \theta_{ada}^1, \theta_{g}^1\}$.
    \item Stage 3: From model $\{\theta_{f}^1, \theta_{ada}^1, \theta_{g}^1\}$, replace $\theta_{g}^1$ with a new generator that can map the embedding space to token space as $\theta_{g'}^0$, update the entire model with data $S_{tgt}$ and a ASR loss such as connectionist temporal classification (CTC), and obtain the final ASR model $\{\theta_{f}^2, \theta_{ada}^2, \theta_{g'}^1\}$.
\end{itemize}
Note that the superscript in each $\theta$ is the number of times the parameters are updated. For example, $\theta_f^2$ means that backbone model has been updated twice, once in stage one and the other in stage three. DRAFT is universal to all self-supervised pretrained models. We verify the effectiveness of DRAFT on E-APC, Bi-APC, Wav2vec2.0 and HuBERT models.

%% file: Tex/exp_steup.tex
\section{Experimental Settings}
\label{sec:exp_setup}
Because of the availability of large databases of adult speech, we explore how SSL methods trained with adult speech can help the development of child ASR systems. In this section, we introduce the data and the experimental settings for the pretraining, adaptation and finetuning stages.

\subsection{Data}
\label{ssec:exp_data}

\subsubsection{Librispeech 960-hour adult speech corpus}
\label{ssec:data_libri}
Librispeech is a widely-used adult speech corpus\cite{panayotov2015librispeech}. It contains 960 hours of read speech extracted from audio books. We use this dataset during the pretraining stage. A 10h subset of the data introduced in \cite{kahn2020libri} is often used for evaluating SSL methods on low-resource tasks, and is referred to as Libri-10h.

\subsubsection{OGI 50-hour child speech corpus}
\label{ssec:data_ogi}
For the fine-tuning experiments, the scripted part of the OGI Kids' Speech Corpus~\cite{shobaki2000ogi} is used. It contains speech from approximately 100 speakers per grade (from kindergarten to grade 10) saying single words, sentences and digit strings. The utterances are randomly split into train (70\%), development (15\%) and test (15\%) sets without speaker overlap. As a result, nearly 50 hours of child data are used to train the child ASR system.

\subsubsection{My Science Tutor (MyST) 240-hour child speech corpus}
\label{ssec:data_myst}
Another corpus used for finetuning is the MyST children speech corpus\cite{ward2011my,ward2019my}. MyST consists of 499 hours with 244,069 utterances of conversational speech between children and a virtual tutor from 1,372 students between third and fifth grades. However, only 42\% of the corpus (240 hours) is annotated for ASR. We use the annotated part of the corpus to verify the effectiveness of our proposed methods. The corpus also contains a development set and test set for evaluation.

\subsection{Pretraining Stage Settings}
\label{ssec:exp_pretraining}
Four self-supervised learning algorithms are investigated: E-APC, Bi-APC, Wav2vec2.0 and HuBERT. All models at this stage are trained with the Librispeech dataset. 

For E-APC and Bi-APC, we use 80-dimensional log-mel filter-bank features ($Z$ in Fig.\ref{fig:ssl_general}) without any concatenation or frame skipping. The features are extracted using a 25ms Hamming window and a frame rate of 10ms. Padding is used for the shorter utterances to make the length be the maximum length of the utterances in a batch. The backbone model $f$ consists of a two-layer convolution block with a sub-sampling of four along the time axis, 12 transformer encoder blocks and a generator for each temporally-shifted sequence. We predict four consecutive frames at each step because of the sub-sampling in the convolution block, resulting in a 320-dimensional output of the generator. Adam optimizer is used with a noam-based scheduler, where the noam factor is 5 and the warmup step is 15k. The model is updated for 130k steps with a batch size of 256. Various starting temporal lags ($s$) and various numbers of consecutive temporally-shifted sequences ($k$) in the E-APC are compared and the best settings in the E-APC are used for the subsequent Bi-APC pretraining.

For Wav2vec2.0 and HuBERT, we directly use the open-sourced pretrained models in the Fairseq\footnote[1]{Our code modified on Fairseq is available at \url{https://github.com/Diamondfan/fairseq}.} toolkit\cite{ott2019fairseq}. We choose the base model that has about 95M parameters to evaluate the effectiveness of the proposed DRAFT framework. Note that the number of parameters in the E-APC and Bi-APC pretraining models are about 39M.

\subsection{Adaptation Stage Settings}
\label{ssec:exp_adapt}
Residual adapters are added to the pretrained model at this stage and only the parameters of residual adapters are updated. We use Xavier uniform initialization \cite{glorot2010understanding} for all RA parameters.

For E-APC and Bi-APC, we adapt the model from the pretraining stage with either the OGI or MyST datasets according to the finetuning task. For the OGI data, residual adapters are updated in 55k steps with a noam factor of 8, warmup steps of 10k. For the MyST data, residual adapters are updated in 74k steps with a noam factor 4 and a warmup step of 15k. The batch size is set to 64 for both datasets.

For Wav2vec2.0 and HuBERT, the residual adapters of Wav2vec2.0/HuBERT are updated in 200k/100k steps with learning rate ramping up from 0 to the peak learning rate in 32k/8k steps, and then decays linearly back to 0, where the peak learning rate is 5e-4. The batch size is set to 16.

We also run experiments using SAFT with the above configurations but all the parameters are updated. Instead, the learning rate is lower than the one used in the pretraining stage (e.g. peaking learning rate of 1e-4 on Wav2vec2.0 and HuBERT). The learning rate is determined empirically after comparisons of various values.

\subsection{Finetuning Stage Settings}
\label{ssec:exp_finetuning}
The CTC loss function is used for the finetuning ASR task training. We train two types of models: 1) a causal transformer with a causal convolution block and encoder blocks with upper-triangular matrices for attention. 2) a non-causal transformer with a regular convolution block and encoder blocks with all-one matrices for attention. The causal transformer is initialized with E-APC pretrained models, while Bi-APC, Wav2vec2.0 and HuBERT use the non-causal transformer. 

In E-APC and Bi-APC, we finetune the model from the pretraining stage or the adaptation stage. The model is updated in 240k steps with a batch size of 32, a noam factor is 2, and a warmup step of 10k steps for the OGI data. For the MyST data, the model is updated in 340k steps with a batch size of 64, a noam factor of 2, and a warmup step of 15k steps. 

In Wav2vec2.0/HuBERT, the model is updated with a batch size of 64 in 40k steps with a multi-step scheduler where the warmup steps are set to 4k. The peak learning rate of 3e-5/7e-5 holds for the next 16k steps, then exponentially decays to the ratio $\lambda$ of the initial learning rate, where $\lambda$ is set to 0.05.

The data augmentation methods, speed perturbation\cite{ko2015audio} and SpecAug\cite{park2019specaugment}, are used for all the experiments at the finetuning stage. Greedy search decoding is used during evaluation. We conduct experiments to decide empirically on the final hyper-parameters. We start the training with a large training epoch (e.g. 100 epochs). Then, if the WER on the validation set does not decrease for several epochs, we stop the training and decide the number of training steps based on the convergence of the training phase.

%% file: Tex/result.tex
\begin{figure*}[t]
\centering
\subfloat[\footnotesize{OGI}]{\includegraphics[width=0.25\textwidth,height=0.25\textwidth]{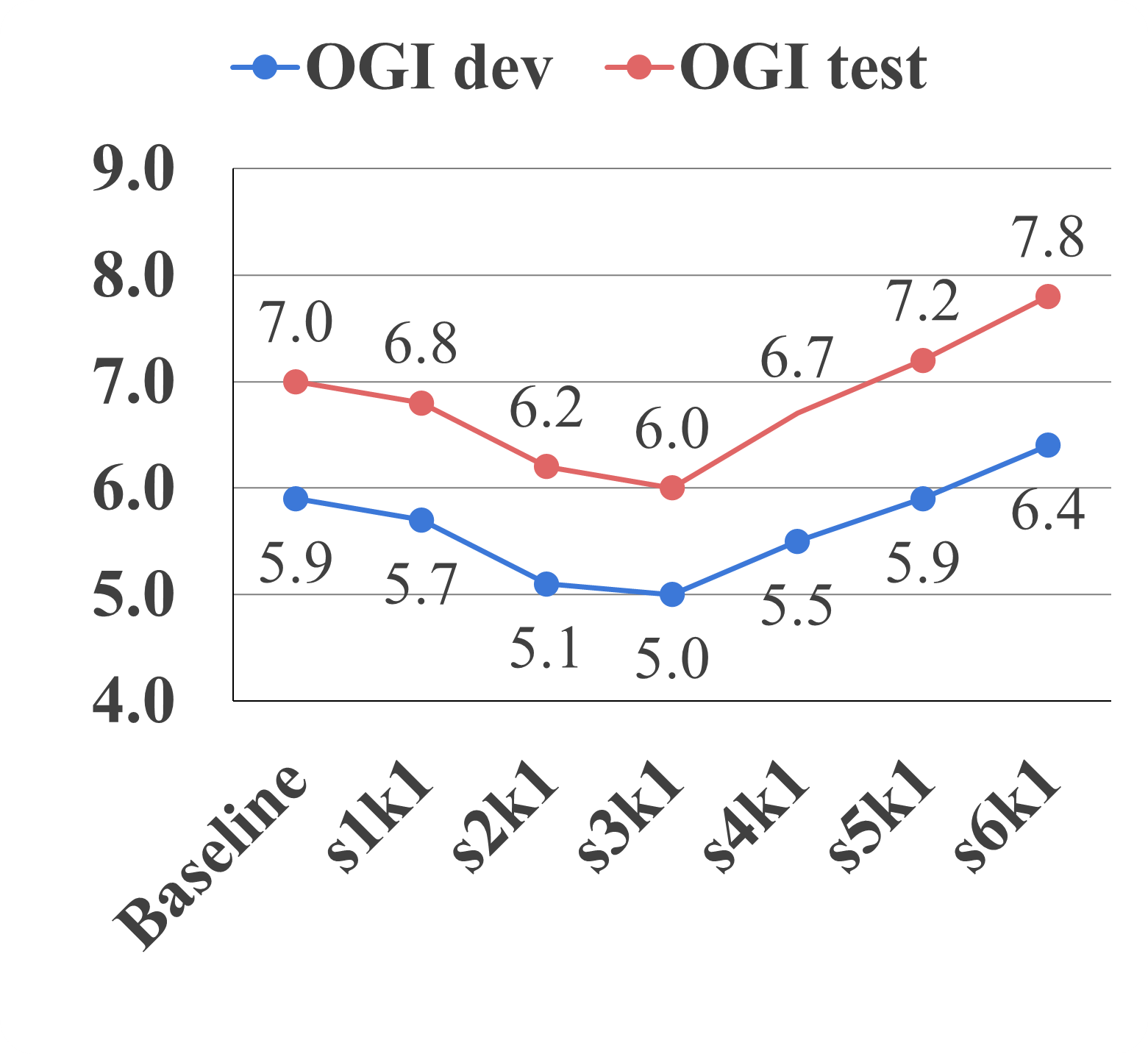}} %
\quad
\subfloat[\footnotesize{MyST}]{\includegraphics[width=0.25\textwidth,height=0.25\textwidth]{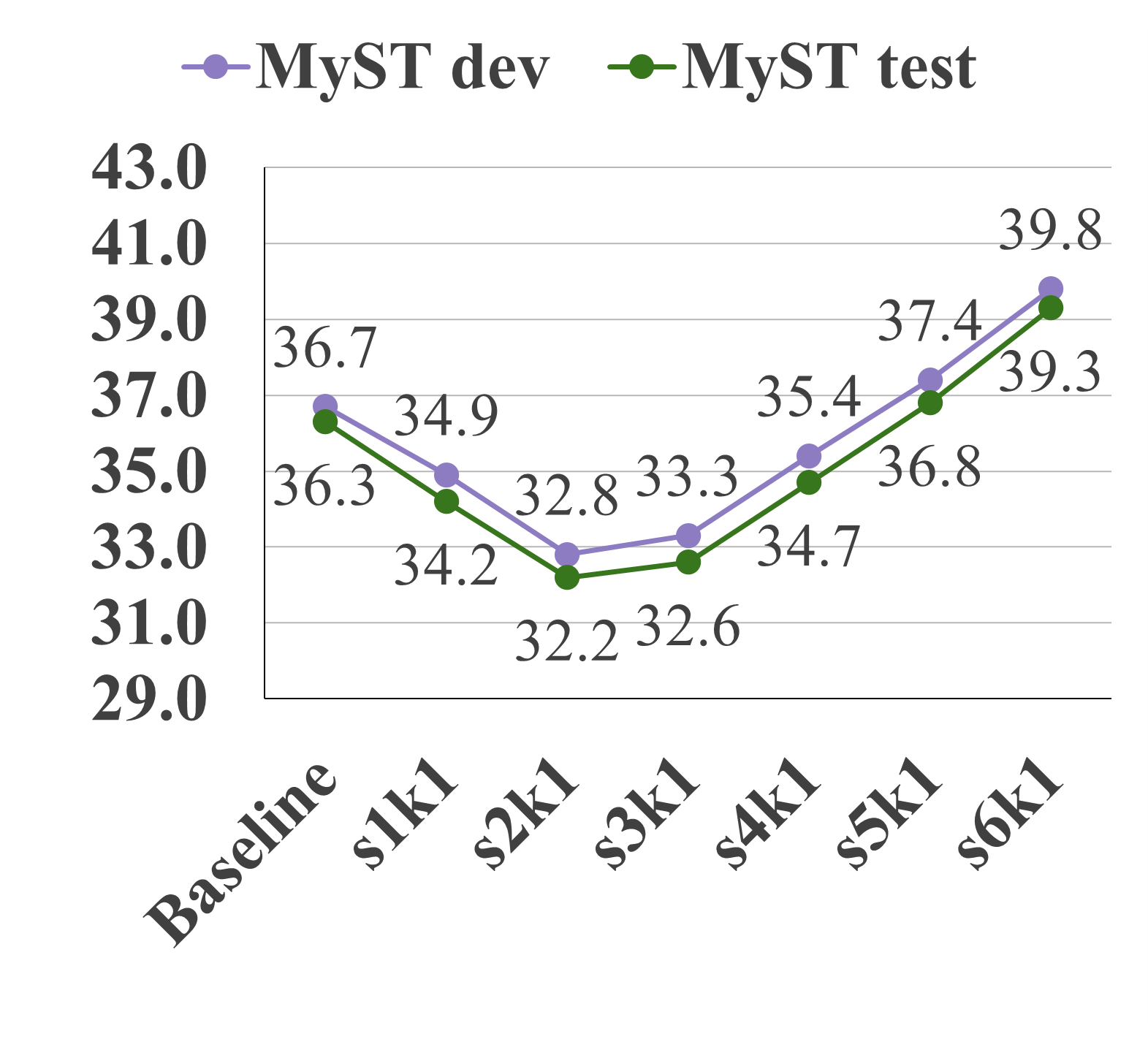}}
\quad
\subfloat[\footnotesize{Libri-10h}]{\includegraphics[width=0.28\textwidth,height=0.25\textwidth]{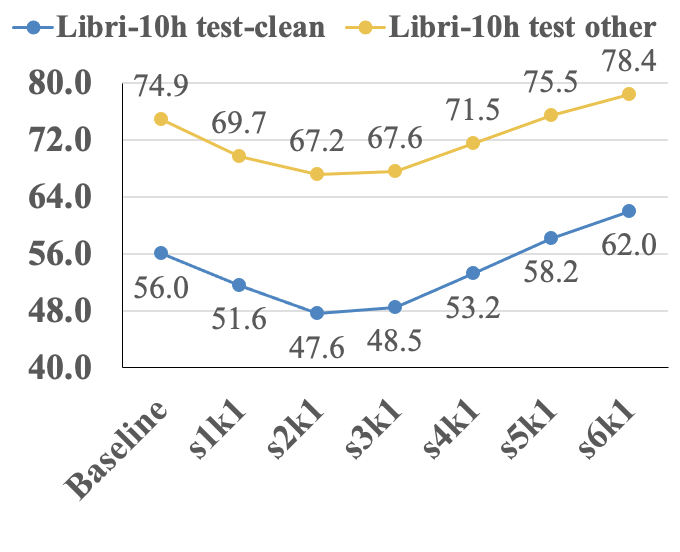}}
\caption{WER results on the OGI, MyST, and Libri-10h datasets for different temporal lags $s$ in APC. Only one temporally-shifted sequence ($k=1$) is used in these experiments. Baseline is the causal transformer trained from scratch. Note that we use adult speech in the pretraining stage. Therefore, (a) and (b) are domain mismatched case, while (c) is the domain matched case.}
\label{fig:apc}
\end{figure*}

\section{Results and Discussion}
\label{sec:result}
The base model (95M) that was pretrained on Librispeech 960 hours data using the Wav2vec2.0 method in the Fairseq toolkit \cite{ott2019fairseq} achieves a WER of 3.4\% on Librispeech test-clean data\cite{baevski2020wav2vec}. However, the model obtains WERs of 41.67\% and 30.77\% for the OGI and MyST test data, respectively, showing a large mismatch between the adult and child speech. In this section, we present experimental results in the same order that we introduced the proposed methods in Sec.\ref{sec:method}. Experiments show how DRAFT alleviates this domain mismatch.  

\subsection{APC for Causal Transformers and Its Extension}
\label{ssec:extension_to_apc}

Fig.\ref{fig:apc} and Table \ref{tab:apc} show ASR results using APC and its extension on the OGI, MyST, and Librispeech 10h\cite{kahn2020libri} datasets. We experimented with the Libri-10h data to show that our methods are not limited to children's speech. The results for the Libri-10h data are similar to those in \cite{kahn2020libri}. The reason for its poor performance compared to finetuning on child speech is because of the amount of finetuning data (10 hours for Librispeech, 50 hours for OGI, and 240 hours for MyST). In addition, OGI is an easier task because of similar distributions for the train and test sets. In Fig.\ref{fig:apc}, we use only one temporally-shifted sequence ($k=1$) and explore the effect of the starting temporal lag $s$. A baseline that does not use any pretraining methods is also included in Fig.\ref{fig:apc}. We can see from the figure that when the prediction lag is more than 4 (16 frames), the WERs increase for all tasks. This result is similar to the results in \cite{chung2020generative} when considering the sub-sampling in the convolution block. The best choice of $s$ is 3 (12 frames) for OGI, and 2 (8 frames) for MyST and Libri-10h. The lags are approximately the duration of an acoustic unit (a vowel or a short syllable). Hence, the model can learn local smoothness of the spectral features by predicting frames within an acoustic unit, and acoustic unit transitions (global structure) by predicting frames in the next acoustic unit, resulting in learning more meaningful speech representations. Then, multiple temporally-shifted sequences are combined to construct E-APC. Based on the results in Fig. \ref{fig:apc}, we experiment with two settings: s1k4, combining four consecutive temporally-shifted sequences starting from 1, and s2k2,  combining two consecutive temporally-shifted sequences starting from 2, because they yielded better results than other settings. The results are shown in Table \ref{tab:apc}. From the table, the performance of $s2k2$ has a $\sim$ 1.8\% relative WER improvement on both tasks compared to the best setting in APC. We also experimented with $L_1$, $L_2$, and a combination of $L_1$ and $L_2$ distance measures as the basic loss function for APC, and found that $L_1$ performs the best.

\input{Tables/table_apc}

\subsection{Bi-APC with A Non-causal Transformer}
\label{ssec:result_biapc}
\input{Tables/table_biapc}
Using the best settings of E-APC ($s2k2$ with $L_1$ distance), we discuss the four solutions of applying Bi-APC to the non-causal transformer mentioned in Sec.\ref{sssec:tsfm_biapc}. Results are shown in Table \ref{tab:biapc}. As can be seen from the table, with more modules shared, the performance tends to be better except that sharing only the generator causes an increase in WERs. However, there is only a 0.1\% absolute WER improvement on the OGI development set when sharing all parameters. The reason may be because the used OGI data (the scripted part) contains shorter utterances (3.5s for OGI, 8.3s for MyST, and 12.8s for Libri-10h), and thus benefit less from Bi-APC. We note a larger improvement with the MyST and Libri-10h data. For example, the WER for MyST test data using the sharing-all solution is decreased from 27.8\% to 25.0\%, and the WER is decrease from 51.9\% to 40.3\% on Librispeech test clean data. The parameters in the separated modules learn very different distributions when they are not shared during Bi-APC pretraining. Averaged parameters may lose information from both sides. As a result, the sharing-all solution outperforms other solutions for Bi-APC pretraining. However, Bi-APC is still worse than methods like Wav2vec2.0 and HuBERT as shown in the next section. When compared to our previous Bi-APC for LSTMs study\cite{fan2021bi}, Bi-APC for transformers performs worse for the child speech databases in terms of relative WERs. We assume that Bi-APC may not be suitable for bidirectional pretraining of models that have only one set of parameters. From this perspective, it is interesting to see whether the transformer can be reformulated as a model that has forward-related and reverse-related parameters.

\subsection{Effect of $d_{ada}$ in DRAFT}
\label{ssec:result_effect_ada}
\input{Tables/table_draft_OGI_MyST}
We conducted experiments with different values of $d_{ada}$ in the residual adapters to examine the impact of the number of adapter parameters, because this number influences both WERs and adaptation efficiency. The experiments are conducted on the OGI and MyST datasets using the E-APC method. Specifically, $d_{ada}$ values are selected from 64 to 2048 and the results are shown in Table \ref{tab:DRAFT_OGI_MyST}. For reference, we also include the results for the baseline, pretraining from E-APC and SAFT. Both WER results and the number of parameters that need to be updated during the adaptation stage are shown in the table. First, when compared to SAFT, DRAFT achieves a better performance with fewer parameters to be updated. In addition, we observe that the WER drops when we increase the number of parameters in the residual adapters. However, the cost is increased training time at the adaptation stage because more parameters need to be updated. We can even achieve an improvement from a WER of 5.9\% to 5.7\% on the OGI test set with only 2\% of the parameters being updated. As a result, the choice of $d_{ada}$ in DRAFT can be adjusted according to different scenarios. For example, one can use a small value of $d_{ada}$ to achieve a fast adaptation of the self-supervised model when computational resources are limited. A large value of $d_{ada}$ can be used to achieve a better performance for the finetuning task. All subsequent DRAFT experiments will use 1024 for $d_{ada}$ since it results in a good trade-off between performance and efficiency.

\subsection{Results of DRAFT with Non-causal Transformers}
\label{ssec:result_draft}
\input{Tables/table_draft_all}
In Sec.\ref{ssec:result_effect_ada}, DRAFT is shown to be effective for E-APC with a causal transformer. Here, we evaluate the DRAFT framework for Bi-APC and other two widely used SSL methods: Wav2vec2.0 and HuBERT. We conduct DRAFT experiments on both the OGI and MyST datasets for E-APC, Bi-APC, Wav2vec2.0 and HuBERT. Results are shown in Table \ref{tab:draft_all}. The table shows that SAFT yields a small improvement or even a negative effect on the WERs compared to the pretraining baselines (without adaptation). The reason may be that updating the entire model causes a catastrophic forgetting of the knowledge learned from adult speech. However, when the proposed DRAFT framework is used, the WERs of the four SSL methods have an improvement on the OGI dataset compared to the pretraining baselines (without adaptation). Specifically, we achieve relative WER improvements of 19.7\%, 3.0\%, 7.4\%, and 16.0\% on the OGI test set for E-APC, Bi-APC, Wav2vec2.0, and HuBERT, respectively. HuBERT achieves the best WER of 2.1\% on the OGI test data. WER improvements are even larger when compared to the baselines without using pretraining methods. For example, we achieve 30\% and 19\% WER improvements on the OGI and MyST data for the E-APC method, respectively. We also observe similar improvements on the MyST dataset, although the relative improvements are smaller than those using the OGI data (child read speech). The reason could be the mismatch in style between MyST data (child spontaneous speech) and pretraining data (adult read speech). Note that we do not try SAFT for Bi-APC because we have already shown its overfitting behaviour in E-APC. For Wav2vec2.0 and HuBERT, the baselines without pretraining are not available because their convergence is difficult to achieve without pretrained models. Note that the results of using HuBERT and Wav2vec2.0 are much better than those of E-APC is because of a better performance of non-causal transformers than causal transformers. However, improving and examining APC with our proposed framework is valuable for causal transformers (streaming models).

Experimental results using the adapter finetuning method \cite{thomas2022efficient} are also presented in Table \ref{tab:draft_all}. Although adapter finetuning has been shown effective in \cite{thomas2022efficient}, it does not perform well in child ASR tasks, maybe because fixing the backbone model (pretrained on adult speech) is not appropriate when the domain mismatch (finetuning on child speech) exists. The results of adapter finetuning also show the importance and effectiveness of DRAFT in reducing domain mismatch. Note that the results of adapter finetuning on Wav2vec2.0 and HuBERT are 100\% because of a slow convergence during training. We may get a reasonable WER for the adapter finetuning method with a longer training schedule. However, the results of adapter finetuning and DRAFT are comparable because they use the same amount of training steps.

\subsection{What do Residual Adapters Learn?}
\input{Tables/table_draft_apc_ogi_detail}
Finally, we explore the behaviour of residual adapters (RA) by using a random initialization of RA or freezing RA parameters during the finetuning stage. By initializing RA parameters randomly ($\theta_{ada}^0$), the WERs in comparison to that of DRAFT which has pretrained RA ($\theta_{ada}^1$), can give us an insight into whether the RA learn the knowledge from the target domain data as expected. Results for E-APC on the OGI data are shown in Table \ref{tab:draft_apc_ogi_detail}. As shown in the table, the performance of DRAFT with learned RA is much better than that when RA are randomly initialized (4.9\% v.s. 5.6\%), showing the successful learning of the target domain knowledge. The table also shows the result when RA parameters are frozen during the finetuning stage. We can see from the table that the performance of freezing RA (5.4\%) is better than the experiment with randomly initialized and updated RA (5.6\%, third row in the table). The results imply that the RA may learn a domain-related transformation from adult to child speech after each block in the transformer. The learned transformation ability from the SSL task might be directly used in the ASR task without further finetuning. One might argue that WER improvements are the results of increasing model parameters. Hence, we conduct an experiment that directly adds residual adapters (``+RA" in the table) in the same way they are added to DRAFT but without any pretrained parameters. The results in Table \ref{tab:draft_apc_ogi_detail} show that DRAFT outperforms the new baseline as well, which could address the concern that the improvements are from increased model capacity.

%% file: Tables/table_apc.tex
\begin{table}[tp]
\centering
\scriptsize
\caption{WER results of APC with $s2k1$ and E-APC with settings of $s1k4$ and $s2k2$. $s1k4$ stands for using four consecutive temporally-shifted sequences starting with a temporal lag of one. A similar meaning applies to $s2k1$ and $s2k2$. $L_p$ norm is the basic loss function used as shown in Eq.\ref{eq:apcloss}. Baseline is the causal transformer trained from scratch.}
\begin{tabular}{l c cc cc cc}
\hline
\multirow{2}{*}{APC} & \multirow{2}{*}{L$_p$}   & \multicolumn{2}{c}{OGI} & \multicolumn{2}{c}{MyST} & \multicolumn{2}{c}{Libri-10h}\\ 
\cmidrule(r){3-4} \cmidrule(r){5-6} \cmidrule(r){7-8} 

& & dev & test & dev & test & clean & other \\ \hline\hline

Baseline & - & 5.9 & 7.0 & 36.7 & 36.3 & 56.0 &	74.9\\ 
APC-s2k1 & L$_1$ & 5.1 & 6.2 & 32.8 & 32.2 & 47.6 & 67.2\\ 
EAPC-s1k4 & L$_1$ & 5.1 & 6.0 & 35.5 & 34.8 & 45.7	& 65.1\\ \hline
\multirow{3}{*}{EAPC-s2k2} & L$_1$ & 5.0 & 6.1 & 32.2 & 31.6 & 45.6 & 65.1\\ 
 & L$_2$ & 5.4 & 6.3 &  32.9 & 32.2 & 48.5 & 67.1\\ 
 & L$_1$ $+$ L$_2$ & 5.2 & 6.3 & 32.4 & 31.7 & 45.6 &	65.1\\ \hline

\label{tab:apc}
\end{tabular}
\end{table}

%% file: Tables/table_biapc.tex
\begin{table}[t]
\caption{WER results of four solutions (shown in Fig.\ref{fig:biapc_trans}) for using Bi-APC for a non-causal transformer. Baseline is the non-causal transformer trained from scratch.}
\scriptsize	
\centering
\begin{tabular}{l ccc cc cc cc}
\hline
\multirow{2}{*}{} & \multicolumn{3}{c}{Share?} & \multicolumn{2}{c}{OGI} & \multicolumn{2}{c}{MyST} & \multicolumn{2}{c}{Libri-10h}\\
\cline{2-10}
& Conv. & Enc. & G. & dev & test & dev & test & clean & other \\
\hline\hline
Baseline & - & - & - & 2.9 & 3.3 & 28.0 & 27.8 & 51.9 & 70.2\\
\hline
\multirow{4}{*}{Bi-APC} 
&  &  &  & 3.2 & 3.9 & 27.8 & 27.3 & 60.7	& 77.0\\
& & & \Checkmark & 3.3 & 4.1 & 32.1 & 31.8 & 58.6 &75.7\\
& & \Checkmark & \Checkmark & 2.8 & 3.4 & 26.2 & 25.7 & 40.1 &58.9 \\
& \Checkmark & \Checkmark & \Checkmark & 2.8 & 3.3 & 25.5 & 25.0 & 40.3 & 58.9\\

\hline

\end{tabular}
\label{tab:biapc}
\end{table}

%% file: Tables/table_draft_OGI_MyST.tex
\begin{table}[t]
\scriptsize	
\centering
  \caption{WER results of different values of $d_{ada}$ in residual adapters. SAFT is the sample adaptation for finetuning that updates the entire model at the adaptation stage. DRAFT is the proposed domain responsible adapter for finetuning that updates only residual adapters at the adaptation stage. The number of updated parameters are also shown in absolute and relative values (compared to the baseline of a causal transformer).}

\begin{tabular}{l c cc cc cc}
\hline
\multirow{2}{*}{~} & \multirow{2}{*}{$d_{ada}$} & \multicolumn{2}{c}{OGI} & \multicolumn{2}{c}{MyST} & \multicolumn{2}{c}{Updated Params} \\ 
\cmidrule(r){3-4} \cmidrule(r){5-6} \cmidrule(r){7-8} 

~ & ~ & dev & test & dev & test & total & relative \\ \hline\hline

Baseline & 0 & 5.9 & 7.0 & 36.7 & 36.3 & 39.2M & 100\%  \\ 

\hspace{2mm}+EAPC & 0 & 5.0 & 6.1 & 32.2 & 31.6 & 39.2M & 100\% \\ \hline

\hspace{4mm}+SAFT & 0 & 5.0& 5.9 & 33.4 & 32.9 & 39.2M & 100\% \\ \hline

\multirow{7}{*}{\hspace{4mm}+DRAFT} 
 & 64 & 4.9& 5.7 & 31.9 & 31.0 & 0.9M & 2\% \\ 
 & 128 & 4.7& 5.6 & 31.6 & 30.9 & 1.7M & 4\% \\ 
  & 256 & 4.6& 5.3 & 31.1 & 30.4 & 3.4M & 9\% \\ 
   & 512 & 4.4& 5.2 & 30.9 & 30.2 &6.8M & 17\% \\ 
 & 1024 & 4.4& 4.9 & 30.1 & 29.4 & 13.7M & 35\% \\ 
& 2048 & 4.4& 4.9 & 30.0& 29.3 & 27.3M & 70\% \\ \hline

\label{tab:DRAFT_OGI_MyST}
\end{tabular}
\end{table}

%% file: Tables/table_draft_all.tex
\begin{table*}[t]
\caption{WER results of using DRAFT for E-APC, Bi-APC, Wav2vec2.0 and HuBERT on the OGI and MyST datasets. Baseline indicates that models are trained from scratch. Improvements of DRAFT are statistically significant ($p<0.05$) compared to the pretraining results. The convergence of Wav2vec2.0 and HuBERT is different to achieve without pretraining.}
\footnotesize
\centering
\begin{tabular}{l  cc cc  cc cc  cc  cc cc cc }
\hline
\rule{0pt}{2ex}
\multirow{3}{*}{} & \multicolumn{4}{c}{E-APC} &
\multicolumn{4}{c}{Bi-APC} & \multicolumn{4}{c}{Wav2vec2.0} & \multicolumn{4}{c}{HuBERT} \\
\cmidrule(r){2-5} \cmidrule(r){6-9} \cmidrule(r){10-13} \cmidrule(r){14-17}  
\rule{0pt}{2ex}
~ & \multicolumn{2}{c}{OGI} & \multicolumn{2}{c}{MyST} & \multicolumn{2}{c}{OGI} & \multicolumn{2}{c}{MyST} & \multicolumn{2}{c}{OGI} & \multicolumn{2}{c}{MyST} & \multicolumn{2}{c}{OGI} & \multicolumn{2}{c}{MyST} \\
\cmidrule(r){2-3} \cmidrule(r){4-5} \cmidrule(r){6-7} \cmidrule(r){8-9} \cmidrule(r){10-11} \cmidrule(r){12-13} \cmidrule(r){14-15} \cmidrule(r){16-17}
~ & dev &  test &  dev & test & dev & test & dev & test & dev & test & dev & test & dev & test & dev & test\\
\hline\hline
Baseline & 5.9 & 7.0 & 36.7 & 36.3 & 2.9 & 3.3 & 28.0 & 27.8 & - & - & - & - & - & - & - & - \\
\hspace{2mm} + Finetune & 5.0 & 6.1 & 32.2 & 31.6 & 2.8 & 3.3 & 25.5 & 25.0 & 2.3 & 2.7 & 17.84 & 17.16 & 2.1 & 2.5 & 17.40 & 16.71 \\
\hspace{2mm} + Adapter Finetune\cite{thomas2022efficient} & 8.6 & 10.1 & 47.4 & 47.3 & - & - & - & - & 100 & 100 & 100 & 100 & 100 & 100 & 100 & 100  \\
\hspace{4mm} + SAFT & 5.0 & 5.9 & 33.4 & 32.9 & - & - & - & - & 2.2 & 2.7 & 17.85 & 17.28 & 2.0 & 2.4 & 17.52 & 16.89 \\
\hspace{4mm} + DRAFT & 4.4 & 4.9 & 30.1 & 29.4 & 2.7 & 3.2 & 24.8 & 24.3 & 2.1 & 2.5 & 17.21	& 16.70 & 1.9 & 2.1 & 16.79	& 16.53 \\
\hline
\end{tabular}
\label{tab:draft_all}
\end{table*}

%% file: Tables/table_draft_apc_ogi_detail.tex
\begin{table}[tp]
\centering
\caption{Experiments showing the behaviour of residual adapters in DRAFT in terms of WER. ``RA Initialization" and ``Update RA?" are describing the finetuning stage. ``+ RA" indicates adding randomly initialized RA at the finetuning stage for a fair comparison to DRAFT. $\theta_{ada}^1$ indicates the RA learned in the adaptation stage and $\theta_{ada}^0$ is the RA with random initialization.}

\begin{tabular}{lcccc}
\hline

\multirow{2}{*}{E-APC} & \multirow{2}{*}{RA Initialization} & \multirow{2}{*}{Update RA?} & \multicolumn{2}{c}{OGI}  \\ \cline{4-5}
~ & ~ & ~ & dev & test \\ \hline\hline
Baseline & None & No & 5.9 & 7.0 \\ 
\noalign{\vskip 1mm}  
\hspace{2mm} + RA & $\theta_{ada}^0$  & Yes & 5.5 & 6.4 \\ 
\noalign{\vskip 1mm} 
\hline
\noalign{\vskip 1mm} 
\multirow{3}{*}{DRAFT} &  $\theta_{ada}^0$ & Yes & 4.8 & 5.6 \\
\noalign{\vskip 1mm} 
~  & $\theta_{ada}^1$ & Yes & 4.4 & 4.9 \\ 
\noalign{\vskip 1mm} 
~ & $\theta_{ada}^1$ & No & 4.7 & 5.4 \\ 
\noalign{\vskip 1mm} 
\hline
\label{tab:draft_apc_ogi_detail}
\end{tabular}
\end{table}

%% file: Tex/conclusion.tex
\section{Conclusions}
\label{sec:conclusion}

In this paper, we developed techniques to improve the performance of self-supervised learning (SSL) methods for children's ASR when unannotated adult speech data are used in the pretraining stage. In the context of autoregressive predictive coding (APC), which is a neural language model style pretraining method for causal transformers, we proposed an extension to APC (E-APC) by learning from multiple temporally-shifted sequences because they contain different levels of information in the structured data. E-APC had a $\sim$ 1.8\% relative WER improvement on the OGI and MyST data compared to APC. In our previous work, a bidirectional APC (Bi-APC) framework was proposed for BLSTM pretraining, which addressed the problem that APC is not suitable for bidirectional model pretraining. In this paper, we further discussed the possibility of using Bi-APC for non-causal transformers. Various solutions were investigated in this paper and results showed that Bi-APC framework can have a slight improvement over the baseline without pretraining, but the results are worse than bidirectional pretraining methods like Wav2vec2.0 and HuBERT. Finally, a domain responsible adaptation and finetuning (DRAFT) framework was proposed to alleviate the domain shifting problem between adult and child speech. The DRAFT framework performed well on E-APC, Bi-APC, Wav2vec2.0 and HuBERT methods, showing that it can improve the performance of pretraining methods for both causal (E-APC) and non-causal transformers  (the other techniques). When compared to the conventional pretraining baselines without adaptation, we achieved relative WER improvements of up to 19.7\% on the two child ASR tasks. The relative WER improvements are even larger (30\% and 19\% for E-APC on the OGI and MyST data, respectively) when compared to the models without using pretraining methods. A future direction of research could be to investigate unsupervised domain adaptation of ASR systems from adult to child speech. In addition, we will explore the pretraining methods with DRAFT for attention-based encoder-decoder models.